\documentclass[prc,twocolumn,floatfix,groupedaddress,showpacs,superscriptaddress] {revtex4}
\usepackage{graphicx}
\usepackage{dcolumn}
\usepackage{mathrsfs}
\usepackage{bm}
\usepackage{dsfont}
\usepackage{dcolumn}
\usepackage[usenames]{color}
\usepackage{amssymb}
\usepackage{amsmath} 
\usepackage{makeidx}
\usepackage{array}
\usepackage{multirow}
\usepackage{bm} 
\usepackage{soul} 
\usepackage[normalem]{ulem}
\DeclareMathAlphabet{\mathpzc}{OT1}{pzc}{m}{it} 

\usepackage{color}

\definecolor{OliveGreen}{rgb}{0,0.6,0}

\begin{document}
\title{Detailed study of decay properties of $^7$He nucleus states in ab initio approach}

\author{D. M. Rodkin}
\affiliation{Dukhov Research Institute for Automatics, 127055, Moscow, Russia}

\affiliation{Moscow Institute of Physics and Technology, 141701 Dolgoprudny, Moscow Region, Russia
}
\author{Yu. M. Tchuvil'sky}
\affiliation{Skobeltsyn Institute of Nuclear Physics, Lomonosov Moscow State University,
119991 Moscow, Russia}
\date{\today}

\begin{abstract}
The energies and decay widths of the states of the exotic $^7$He nucleus are studied  in an ab initio approach. The spectrum of these states is calculated using No-Core Shell Model and corresponding extrapolation procedure. Well-proven on a large amount of data the Daejeon16 potential is used in the calculations. The previously developed NCSM-based approach, which includes a method for constructing the basis of orthogonalized functions of cluster channels and a procedure for matching the cluster form factors obtained within this method with the corresponding asymptotic wave functions, is applied to compute the decay widths of the  levels. The possibilities of the approach for  calculating the partial decay widths of nuclear states into various channels strongly differing in type of fragmentation, spin, angular momentum of relative motion, and amplitude are demonstrated. The results obtained are compared with the results of other microscopic calculations.
\end{abstract}

\pacs{}
\maketitle

\section{Introduction}
In recent decades, research related to low-energy nuclear physics has shown two basic trends.

In experimental research, there has been a rapid transition from the study of stable and neighboring long-lived nuclei to the study of exotic, including nucleon-unstable systems. The most important factor in the development of such  a research is the upgrading of existing and constructing of new radioactive beam facilities. Programs for upgrading existing facilities are being implemented in RIKEN (creation of RIBF), GANIL (SPIRAL2), GSI (FAIR), NSCL (FRIB). Building of new centres for the study of radioactive isotopes: HIAF and RAON is in progress.

In theoretical  research, largely due to the rapid progress of computer technologies, an increasing place is occupied by high-precision microscopic approaches, in particular ab initio (from first principles) methods of describing nuclear systems.

An important place among ab initio methods is taken by: different versions of No-Core Shell Model (NCSM) \cite{ncsm1, ncsm2, ncsm3, dyt1, dr}, Gamov Shell Model (GSM) \cite{gamov1, gamov2, gamov3}, Green's functions Monte Carlo Method \cite{gfmc1,
gfmc2, gfmc3, gfmc4}, the Coupled Cluster Method \cite{ccm1, ccm2}, Hyperspherical Harmonic Approach \cite{hsh1, hsh2}, In-Medium  Similarity Renormalization Group \cite {imsrg}, Self Consistent Green's Function Method \cite {scgf1, scgf2, scgf3}, and Lattice Effective Field
Theory for Multi-nucleon Systems \cite{left1, left2}. These methods are based
on realistic two-nucleon (NN) and three-nucleon (NNN) potentials. These potentials could be derived from Chiral
Effective Field Theory \cite{dj16, machleidt1, machleidt2} or from nucleon
scattering data by the use of $J$-matrix inverse scattering method \cite{jisp1}.

At present, there is a process of a sharp increase in the attention of various theoretical groups to the problems of describing unbound nuclear states within the framework of ab initio approaches or methods approaching them in terms of their theoretical level. This trend is clearly reflected in review \cite{johns}.

In view of the mentioned above, the choice of the object of research and the methodology for solving the problem posed in this work seems to be timely.

It should be noted that ab initio approaches focused on the discussed problem are already presented in the literature.
Among them the methods which combine NCSM and Resonating Group Model (RGM) namely No-Core Shell
Model / Resonating Group Model (NCSM/RGM) \cite{ncsmrg1} and No-Core Shell Model with
Continuum (NCSMC) \cite{ncsmc0,ncsmc1, ncsmc2, ncsmc3, ncsmc4, ncsmc5, ncsmc6} seem to be the most versatile.  To describe resonances (including those undergoing multiparticle decay) using calculations of scattering phase shifts the NCSM-based SS-HORSE method was proposed \cite{sshorse1,sshorse2}. Some nuclear resonances can be also studied with the use
of mentioned  above GSM \cite{gamov3}. 

A significant place among the methods of high-precision description of unbound states of nuclei is occupied by ab initio microscopic approaches, which are not, in strict definition, ab initio, but accurately take into account almost all dynamic properties of unbound systems. Such approaches are, in particular, the Microscopic Cluster Model (MDM) \cite{desc1,desc2,desc3}, the version of RGM \cite{kv} which exploits realistic NN-potentials, and the Fermionic Molecular Dynamics (FMD) \cite{neff1,neff2,neff3}.  Admittedly, all mentioned above approaches aimed at describing unbound states can be applied to a very limited number of nuclear resonances compare to the list of bound  states whose total binding energies (TBEs) and electromagnetic properties are  described by conventional ab initio methods, NCSM, for example. 

In our previous papers \cite{our4,our5,our6}, an ab initio approach was developed. It allows one to calculate the asymptotic characteristics of the real decay of unbound and virtual splitting of bound states of nuclei namely the total and partial decay widths and the asymptotic normalization coefficients (ANCs) of certain channels, respectively. This approach includes NCSM as a basic building block. 

Both in the previous works and in the present one, at the first stage the A-nucleon Schr\"odinger equation  with realistic NN-potentials is solved using the basis of totally antisymmetric A-nucleon oscillator wave functions (WFs). M-scheme, in which one-nucleon functions are characterized by the angular momentum $j_i$ and its projection $m_i$, is exploited. The discussed basis is built to be complete up to the maximal total number of oscillator quanta $N^{max}_{tot}$. The size of the used basis in the M-scheme in our calculations is usually in the range $10^{8} \div 10^{9}$ components. The binding energies and the WFs of the ground, excited, and resonance states of nuclei are computed. In the current paper we use well-proven in calculating the spectra of nuclei with mass A$\leq$16, their sizes and other characteristics the Daejeon16 potential \cite{dj16}, which is based on Chiral Effective Field Theory.  
Next, so-called Cluster Channel Orthogonalized Functions Method (CCOFM) \cite{our1, our2,our3} is applied.  The procedure, as a whole, looks as follows. A basis of the orthogonalized cluster-channel WFs is built by this method, and the eigenfunctions of using Hamiltonian computed in the NCSM are projected onto the functions of this basis. At the third stage, the functions obtained within the projecting procedure  -- so called cluster form factors (CFFs) -- are matched with the asymptotic WFs of the corresponding channels.

The most important problem determining the prospects for the application of the discussed approach as a whole is that the range of distances where solutions of the Schr\"odinger equation are correctly described by NCSM  A-nucleon WFs expands proportionally to the size of the classically allowed area of the harmonic-oscillator potential and therefore proportionally to  $[N_{max}^{tot}]^{1/2}$. So this range is somewhat limited.   Does the CFF obtained in these calculations reproduce the shape of the asymptotic WF at inter-cluster distances where the nuclear interaction is negligible?
The results of our studies presented in Refs. \cite{our5,our6} give the answer yes to this question for nuclei $^7$Li and $^8$Be which are close to $^7$He in mass, and not only for one-nucleon, but even for cluster channels. 

We also point out that the methodological scheme of this work includes a new element in comparison with those discussed above. For a more precise determination of the energies of the levels calculated in NCSM, the extrapolation procedure presented in Ref. \cite {extrapA} is used.

In Sections II and III, we give a brief outline of the just described methodology.

In this paper we present  the results of its application to the detail theoretical study of characteristics of $^7$He nucleus spectrum. The choice of this object is determined by the following motivations. Despite the fact that the $^7$He system has no bound states, there are experimental data on some of its states. The prospects for a more detailed experimental study of this object appear to be good. We have successful experience in studying nuclei of the same and close mass.The literature contains theoretical works devoted to ab initio studies of this object, which makes it possible to compare the results of the approaches used in them with the results of this one.

\section{Outline of methodology of the approach}

\subsection{Total binding energy extrapolation procedure}

In our previous works \cite{our4,our5,our6}, it was shown that even a not too large deviation of the calculated resonance energy from the experimental one leads to a noticeable change in the calculated value of the decay width. For this reason, the use of experimental resonance energies in calculating decay widths is preferable.
In the framework of theoretical study of exotic nuclei such as $^7$He, it is much more difficult to rely on experimental results due to their incompleteness, unreliability or absence. So, the requirements for the accuracy of calculating the level energies  for this kind of problems are stringent. At present, despite broad capabilities of the shell model calculations, for neutron-rich nuclei such as $^6$He and $^7$He they are not fully converged. The only way out of this situation is the use of one of the extrapolation procedures. There are three widely known techniques, namely: deep learning extrapolation tool based on the artificial neural networks (ANN) \cite{extrap}, or five-parameter "Extrapolation A5" \cite{extrapA}. or three-parameter "Extrapolation B" \cite{extrapB} methods. The "Extrapolation B" method adopts a three-parameter extrapolation function that contains a term that is exponential in total number of quanta above the minimum harmonic oscillator energy configuration (cut-off parameter) $N^{*max}_{tot}$. The "Extrapolation A5" method adopts a five-parameter extrapolation  function that contains a term proportional to ($exp(-{N^{*max}_{tot}}^{1/2})$ in addition to the single exponential term $exp (- N^{*max}_{tot}$) used in the "Extrapolation B" method. According to paper \cite{extrap} both these methods give comparable results with ANN deep learning tool, at the same time they are much easier. "Extrapolation A5" method demonstrates slightly smaller deviation values, so in our work we used this method for obtaining total binding energies of $^7$He and $^6$He states in "infinite" shell-model basis. The extrapolation function proposed in Ref.  \cite{extrapA} depends on the five free parameters: $E_{\infty}, a, c, d, k_{\infty}$ and also on parameters $b = \sqrt{{\hbar/}{m \omega}}$, $\Lambda_{i} = b^{-1}\sqrt{2(N^{*max}_{tot,i} + 3/2)}$, $L_{i} = b\sqrt{2 (N^{*max}_{tot,i} + 3/2)}$, $L_t = L_i + 0.54437b(L_{i=0}/ b)^{-1/3}$. It is written in the following form:
   
\begin{equation}
E_{state}(N^{*max}_{tot,i}, \hbar \omega) = E_{\infty} + a \cdot exp(-c \Lambda_{i}^2) + d \cdot exp(-2 k_{\infty} L_t) \label{extr}
\end{equation}

The values of free parameters are determined for each level independently by fitting this function  to theoretically calculated total binding energies. For the $^6$He nucleus states we fit these parameters  in the $\hbar \omega$ range from 10 to 25 MeV with 2.5 MeV step and $N^{*max}_{tot,i}$ = 4, 6, 8, 10, 12, 14. For the $^7$He nucleus states the $\hbar \omega$ range is from 2.5 MeV to 22.5 MeV with 2.5 MeV step and $N^{*max}_{tot,i}$ = 5, 7, 9, 11, 13. Changes in energy values resulting from this fit are presented  bellow in tables I -- III.

\subsection{Cluster channel orthogonalized basis constructing}

The method of construction of the basis of the WFs of cluster channels is described in detail in our previous works \cite{our1,our2,our3,our4,our5,our6}, so here we demonstrate only its key points.

Let us consider translationally invariant A-nucleon WFs of arbitrary two-fragment decay channel basis 
corresponding to the separation A = A$_1$ + A$_2$ 
\begin{equation}
\Psi^{\{k_1,k_2\}} _{{A\,},nlm}  = \hat A\{\Psi^{\{k_1\}} _{A\,_1} \Psi^{\{k_2\}}
_{A\,_2} \varphi _{nlm} (\vec \rho )\} _{J_c,M_{J_c},M_JT},  \label{eq1}
\end{equation}
where  $\hat A$ is the antisymmetrizer, $\Psi^{\{k_i\}}
_{A\,_i}$ is a translationally invariant internal WF of the fragment labelled by a set
of quantum numbers $\{k_i\}$; $\varphi _{nlm} (\vec \rho )$ is the WF of the relative
motion. The channel WF as a whole is labelled by the set of quantum numbers $c_\kappa$ that
includes $\{k_1\},\{k_2\},l,J_c,M_{J_c},M_J,T$, where $J_c$ is
the channel spin. The basic idea of the method is to represent each function of the cluster basis
as a linear combination of the functions of the M-scheme.  To do that function (\ref{eq1})
is multiplied by the function of the center of mass (CM) zero vibrations $\Phi _{000}
(\vec R)$. Then the transformation of WFs caused by changing from $\vec R,\vec \rho$ to
$\vec R_1 ,\vec R_2$ coordinates -- different-mass Talmi-Moshinsky-Smirnov transformation \cite{talmi-moshinsky} --
 is performed and WF (\ref{eq1}) takes the form

\begin{equation}
\begin{array}{rcl}
\Phi _{000} (\vec R)\Psi^{\{k_1,k_2\}} _{{A\,},nlm} = \sum\limits_{N_i ,L_i ,M_i}
{\left\langle {{\begin{array}{*{20}c}
   {000}  \\
   {nlm}  \\
\end{array}  }}
 \mathrel{\left | {\vphantom {{\begin{array}{*{20}c}
   {000}  \\
   {nlm}  \\
\end{array}  } {\begin{array}{*{20}c}
   {N_1 ,L_1 ,M_1 }  \\
   {N_2 ,L_2 ,M_2 }  \\
\end{array}}}}
 \right. \kern-\nulldelimiterspace}
 {{\begin{array}{*{20}c}
   {N_1 ,L_1 ,M_1 }  \\
   {N_2 ,L_2 ,M_2 }  \\
\end{array}}} \right\rangle } \\[\bigskipamount]
\hat A\{ \Phi _{N_1 ,L_1 ,M_1 }^{A_1 } (\vec R_1 ) \Psi^{\{k_1\}}_{A\,_1 } \Phi _{N_2
,L_2 ,M_2 }^{A_2 } (\vec R_2 )\Psi^{\{k_2\}} _{A\,_2 } \} _{J_c,M_{J_c},M_JT} . \label{eq2}
\end{array}
\end{equation}

 The key technical procedure of the method is to transform each of the two products of the internal WF of the fragment by the function of non-zero oscillations of its CM into a superposition of Slater determinants (SDs)

\begin{equation}
\Phi _{N_i ,L_i ,M_i }^{A_i } (\vec R_i )\Psi^{\{k_i\}}_{A\,_i }  = \sum\limits_k
{X_{N_i ,L_i ,M_i }^{A_i (k)} \Psi _{A\,_i (k)}^{SD} }. \label{eq3}
\end{equation}
It is the possibility of implementing this procedure that imposes restrictions on the list of cluster channels available for research within the framework of this method. Quantity $X_{N_i ,L_i ,M_i }^{A_i (k)}$ is called a cluster coefficient (CC). Technique of
these objects is presented in detail in \cite{nem}. Various approaches to calculating these coefficients, convenient in various special cases, are presented in papers \cite{ar,sa1,sa2,sa3,sa4} In the present work we use the formalism based on the method
of the second quantization of the oscillator quanta described in details in \cite{our5,our6}. As a result of these transformations, the antisymmetrized product of functions in the right-hand side of expression (\ref{eq2}) also turns out to be a superposition of SDs.

It should be noted that WFs of cluster-channel basis terms (\ref{eq1}) of one and the same
channel $c_\kappa$ characterized by the pair of internal functions
$\Psi^{\{k_1\}}_{A_1}$, $\Psi^{\{k_2\}}_{A_2}$  and extra quantum numbers
$l,J_c,J,M_J,T$ (i. e. the vector coupling is meant here) briefly denoted as $\Psi^{c_\kappa} _{{A\,},n}$ are non-normalized due to the properties of the antisymmetrization operator and, with rare exceptions, non-orthogonal. Creation of orthonormalized basis functions of a separate channel $c_\kappa$ is performed by the diagonalization of the overlap kernel matrix

$$ ||N_{nn'} || \equiv  \langle \Psi^{c_\kappa} _{{A\,},n'}|\Psi^{c_\kappa} _{{A\,},n} \rangle = $$
\begin{equation}
\langle\Phi _{00}(\vec R) \Psi^{\{k_1\}} _{A_1} \,\Psi^{\{k_2\}} _{A_2} \,\varphi _{nl} (\rho )
|\hat A^2 |\Psi^{\{k_1\}} _{A_1} \,\Psi^{\{k_2\}} _{A_2} \,\varphi _{n'l} (\rho )\Phi _{00}(\vec R) \rangle . \label{eq7}
\end{equation}

The eigenvalues and eigenvectors of this overlap kernel are the same in the shell-model and translationally-invariant representations and can be written as:

\begin{equation}
\varepsilon _{\kappa,k}  =  \langle \hat A\{ \Psi^{\{k_1\}} _{A_1} \,\Psi^{\{k_2\}} _{A_2} \,f_l^k (\rho
) \} |\hat 1|\hat A\{ \Psi^{\{k_1\}} _{A_1} \,\Psi^{\{k_2\}}_{A_2} \,f_l^k (\rho
) \}  \rangle ; \label{eq8}
\end{equation}

\begin{equation}
f_l^k (\rho ) = \sum\limits_n {B_{nl}^k \varphi _{nl} (\rho )}. \label{eq9}
\end{equation}

On the other hand, the WFs of the orthonormalized channel basis $c_\kappa$ 

\begin{equation}
\Psi^{SD,c_\kappa} _{{A\,},kl}=\varepsilon^{-1/2} _{\kappa,k}|\Phi _{00}(\vec R)\hat A\{ \Psi^{\{k_1\}}
_{A_1} \,\Psi^{\{k_2\}}_{A_2} \,f_l^k (\rho ) \}  \rangle. \label{eq10}
\end{equation}
turn out to be represented in the form of the superposition of the SDs. The basis of such functions is complete in the sense that a function of this channel

\begin{equation}
\Psi^{SD,c_\kappa} = \Phi _{00}(\vec R)\hat A\{\Psi^{\{k_1\}} _{A\,_1} \Psi^{\{k_2\}}
_{A\,_2}\Phi(\rho) Y_{l} (\Omega)\} _{J_c,M_JT}  \label{eq01}
\end{equation}
including arbitrary WF $\Phi(\rho)$ can be represented as a superposition of such WFs. 

These functions can be introduced into the basis of the NCSM calculations for its expansion towards large  $N^{tot}_{max}$ or used to find the cluster characteristics of nuclear states calculated in the NCSM basis, both extended due to their inclusion, and the traditional one.

\subsection{Cluster characteristics of in ab initio calculations}

In this work, the energies of the levels of the initial and resulting nuclear states are calculated in the usual M-scheme, and the basis described above is used to calculate the cluster characteristics of decay channels: the cluster form factor and the spectroscopic factor (SF).

The CFF $\Phi^{c_\kappa}_A(\rho)$  describes the relative motion of subsystems in A-nucleon configuration space and is defined by the following overlap
$$\Phi^{c_\kappa}_A(\rho)=$$
\begin{equation}
\langle \Psi _{A}|\hat N^{-1/2}\hat A\{\Psi^{\{k_1\}} _{A\,_1} \Psi^{\{k_2\}}
_{A\,_2}\frac{\delta(\rho-\rho')}{\rho'^2}Y_{l} (\Omega)\} _{J_c,M_JT}\rangle ,\label{cff}
\end{equation} 
where $\Psi _{A}$  is the WF of the initial nucleus -- the  NCSM solution of the A-nucleon Schr\"odinger equation, and $\hat N$ is the norm operator of the generalized function of the cluster channel which takes the form:

$$\hat N=\langle \hat A\{\Psi^{\{k_1\}} _{A\,_1} \Psi^{\{k_2\}}
_{A\,_2}\frac{\delta(\rho-\rho')}{\rho'^2}Y_{l} (\Omega)\} _{J_c,M_JT}|\times$$
\begin{equation}
|\hat A\{\Psi^{\{k_1\}} _{A\,_1} \Psi^{\{k_2\}}
_{A\,_2}\frac{\delta(\rho-\rho')}{\rho'^2}Y_{l} (\Omega)\} _{J_c,M_JT}\rangle.\label{norm}
\end{equation} 
Representation of the generalized function of the relative motion in the form of an expansion in terms of oscillator functions

\begin{equation}
[\delta(\rho-\rho')/\rho'^2]Y_{lm} (\Omega)=\sum\limits_n\varphi _{nlm} (\vec \rho )\varphi _{nlm} (\vec \rho' )\label{delta}
\end{equation}
first, reduces the norm operator to the overlap kernel matrix (\ref{eq7}) and, second, makes it possible to write the CFF in the form 
 
\begin{equation}
\Phi^{c_\kappa}_A(\rho)=\sum\limits_k \varepsilon^{-1/2} _{\kappa,k} \langle \Psi _{A}|\hat A\{ \Psi^{\{k_1\}} _{A_1}
\,\Psi^{\{k_2\}}_{A_2} \,f_l^k (\rho') \} \rangle f_l^k (\rho ).
\end{equation}
 
 After that the CCF can be expressed in the form of an expansion in the oscillator basis using above presented techniques:
\begin{equation}
\Phi^{c_\kappa}_A(\rho)=\sum\limits_k\varepsilon^{-1/2} _{\kappa,k}\sum\limits_{n, n'}B_{nl}^k B_{n'l}^{k}
C_{AA_1A_2}^{n'l}\varphi _{nl} (\rho ) \label{eq11}
\end{equation}
 The coefficient contained in this expression has the form
$$C_{AA_1A_2}^{nl} = \langle \hat A\{\Psi^{\{k_1\}} _{A_1} \Psi^{\{k_2\}} _{A_2} \varphi _{nl}
(\rho ) \} | \Psi _{A} \rangle =$$
\begin{equation}
\langle \Psi^{SD,c_\kappa} _{A, nl}|\Phi _{000} (R)| \Psi _{A} \rangle = \langle \Psi^{SD,c_\kappa} _{A,
nl}|\Psi^{SM} _{A} \rangle. \label{eq12}
\end{equation}
This coefficient is traditionally called the spectroscopic amplitude (SA). A number of very diverse methods of its
calculation depending on the masses of the initial nuclei and fragments were described
in \cite{nem,sa1, sa2, sa3}. All of them are based on the CCs formalism.

The SF is defined as the norm of CFF, for the discussed channel $c_\kappa$it can be written as
$$S_l^{c_\kappa}  = \int {|\Phi^{c_\kappa}_A(\rho)|^2 } \rho ^2 d\rho =$$
\begin{equation}
\sum\limits_k {\varepsilon _k^{ - 1} \sum\limits_{nn'} {C_{AA_1A_2}^{nl} }
C_{AA_1A_2}^{n'l} B_{nl}^k } \;B_{n'l}^k . \label{eq14}
\end{equation}

 It should be noted that an alternative method for calculating CFFs (and obviously SF, if necessary) that does not use the formalism of cluster coefficients is presented in the literature \cite{cffn}.

The definitions of the CFF (\ref{cff}) and SF (\ref{eq14}) are completely
equivalent to those proposed in \cite{newsf1} (the so-called “new” spectroscopic
factor and CFF as opposed to "old" ones). In contrast to the traditional definition, the new
 CFF and SF characterize the total contribution of orthonormalized cluster
components to the solution of the Schr\"odinger equation  describing an A-nucleon system. Reasons for the
necessity of its use to describe decays and reactions can be found in \cite{newsf2,
newsf3}. In \cite{vt1, vt2}, it was shown that the correct definition eliminates
a sharp contradiction between  theoretically calculated values of the cross sections for
reactions of knock-out and transfer of $\alpha$ clusters and experimental data. 

Here it is pertinent to make an essential remark. For the decay of the $^7$He nucleus discussed here, only neutron channels are relevant. In contrast to cluster channels, the "old" definition is still often used for form factors (FFs), SAs (\ref{eq11}), and SFs of single-nucleon channels. This definition ignores the necessary renormalization of  any, including one-nucleon FF by the norm kernel. This is due to a well-established tradition and the fact that the numerical differences in the results of calculations in the "old" and "new" definitions are usually not large for one-nucleon channels. We have to stress that only ``new'' version is used in the current work and we consider this to be the only correct approach.

\subsection{Application of CFFs for studies of decay  characteristics of nuclear states}
 
Both SFs and CFFs are the objects used in theoretical studies on nuclear decays and reactions. Evidently CFF is more informative characteristic of a cluster channel. In the current work the obtained CFFs are exploited for computing the widths of resonances. The norms of these values  -- SFs -- are used to distinguish the main channels against the background of a multitude of other ones, existence of which practically do not affect the results of experiments.  

As in our previous works \cite{our5,our6}, we use the procedure of matching the CFF with the asymptotic WF of the corresponding channel. 
The results of our studies presented in these works demonstrate that the CFF in its new definition allows matching with the asymptotic WF at relatively small distances, where the nuclear interaction is negligibly weak, but exchange effects
generated by the antisymmetry of the total channel WF and manifested through the exchange terms of the overlap kernel are still not negligible. "Switch off" the effects of antisymmetrization -- the vanishing of the matrix elements of the permutation operators included in the antisymmetrizer -- occurs at larger distances. This is precisely what constitutes the advantages of the "new" definition as applied to the study of nuclear decays. This property is very important for dealing with NCSM CFFs.

So in the discussed approach, a direct matching procedure described in classical textbooks of quantum mechanics  is applied to calculate the widths of narrow resonances. For such resonances or, more precisely, for those of them whose small width is due to a
low penetrability of the potential barrier, we used a very compact procedure proposed in
\cite{matchingpoint}. This procedure is applicable because  for such resonances there is  sufficiently
wide range of distances in which the nuclear attraction is already switched off and at the same time 
the potential barrier is high enough.
At any inner point $\rho_{in}$ of this area, the relationship between the regular and irregular solutions 
of the two-body Schr\"odinger equation in the Wentzel-Kramers-Brillouin (WKB) approximation has the form

\begin{equation}
F_l (\rho_{in} )/ G_l (\rho_{in} ) = P_l (\rho_{in} ) \ll 1,\label{eq15}
\end{equation}
where $P_l (\rho_{in} )$ is the penetrability of the part of the potential barrier that
is located between the point $\rho_{in}$ and the outer turning point. The smallness of
this penetrability is the condition of applicability of the approximation in which the
contribution of the regular solution can be neglected.
To determine the position of the matching point $\rho_{m}$ of the CFF and irregular WF  in this
range, we use the condition of equality of the logarithmic derivatives

\begin{equation}
\frac{d\Phi _A^{c_\kappa  } (\rho)/d\rho}{\Phi _A^{c_\kappa  } (\rho )}= \frac{dG_l
(\rho)/d\rho}{{G_l (\rho )}}. \label{eq16}
\end{equation}
Comparison of the values of the CFF and function  $G_l (\rho)$ in the matching point allows one to determine the amplitude of the channel WF in the asymptotic region which takes the form $\beta G_l (\rho)$,  where
\begin{equation}
\beta = \Phi _A^{c_\kappa  } (\rho
_{m})/G_l (\rho _{m}). \label{ass}
\end{equation}

As a result, the decay width is given by the expression
\begin{equation}
\Gamma  = \frac{{\hbar ^2 }}{{\mu k}}\left [ \frac{{\Phi _A^{c_\kappa  } (\rho
_{m})}}{{G_l (\rho _{m})}}\right ]^2. \label{eq17}
\end{equation}

To make the list of the properties  of the states of a certain nucleus broader,  large-width
resonances are considered too. If the resonance is wide and so the penetrability $P_l
(\rho_{in} )$ (\ref{eq15}) is not small the width of this resonance is calculated using the
 simple version of the conventional R-matrix theory:

\begin{equation}
\Gamma  = \frac{\hbar^2}{\mu k_0} (F^2_l ( \rho_{m} ) + G^2_l ( \rho_{m} ))^{-1}
(\Phi _A^{c_\kappa} (\rho_{m} ))^2. \label{eq20}
\end{equation}

Naturally the use of this version leads to reduction in accuracy of calculation results. However, this technique seems natural for describing experimental decay widths, since when extracting their values from the cross sections of resonance processes, different versions of the R-matrix theory of nuclear reactions are also used. Besides that, the accuracy of the data, concerning large-width resonances,  both decay widths and excitation energies, extracted from various experiments, is also very limited. It is important to note that, in contrast to calculating the total fragmentation width of any resonant state, finding the partial widths of its decay into various channels requires calculating the amplitudes of channel WFs in the asymptotic region.

In this paper the proposed approach is utilized to study the spectrum of resonance states of the $^7$He nucleus and the partial decay widths of these states.

An important point is that the Daejeon16 potential \cite{dj16} is exploited as a model of NN-interaction in the current work. 
It is built using the N3LO limitation of Chiral Effective Field Theory \cite{ceft1} softened by Similarity
Renormalization Group (SRG) transformation \cite{srg1}. This potential is designed to calculate 
all kinds of characteristics of nuclei with the masses $A \leq 16.$ It was tested in the framework of large-scale computations of the
total binding energies, excitation energies, radii,  moments of nuclear states and the reduced probabilities of electromagnetic transitions. 
These tests demonstrated that such characteristics are, in general, reproduced well.  Besides that this choice is supported by our previous studies of the asymptotic characteristics of the cluster channels of light nuclei, in which other NN-potentials were also involved in the analysis. The results of these studies are presented in the Refs. \cite{our1, our3, our4}.

\section{Results of the calculations and discussion}

In this work, we calculate the total binding energies (TBEs), excitation energies, as well as decay energies and widths of the levels of $^7$He system; and also calculate TBEs of the lower levels of the fragment $^6$He nucleus necessary for that. The NCSM calculations were carried out with the use of newest version of Bigstick shell model code \cite{bigstick}. The basis is limited by the value of cut-off parameter $N^{*max}_{tot}$=13 i. e. maximal total number of oscillator quanta N$^{max}_{tot}$ =16. For a limited basis, the optimal value of the oscillator parameter turns out to be $\hbar\omega$=12.5 MeV.

Let us consider, first of all, the calculated values of the total binding energy and the excitation energy of the $^6$He nucleus. They are presented in Tab. I. Despite the fact that the ground state of the $^6$He nucleus is a classic example of a nuclear system with a two-neutron halo, the experimental value of its TBE ($E_{exp.}$) is well reproduced in the shell-model calculations on a limited basis (values of such a type are denoted by symbol $E^{lim.}_{th.}$). The extrapolated TBE value $E^{extr.}_{th.}$ deviates from the experimental result even somewhat more significantly, but, in any case, the deviation does not exceed 0.4\%. The magnitude of the absolute deviation which is equal to 128 keV also appears to be satisfactory. This is not the case for the first excited state, for which the deviation of the TBE calculation result from the experimental one is 273 keV. This fact, and especially the significant overestimation of the excitation energy for this level, demonstrates the need to use an extrapolation procedure. As for the second excited state of the  $^6$He nucleus, its gigantic width ($\sim$ 12 MeV), obviously, does not allow simultaneously evaluating its excitation energy from the experiment with an accuracy better than several MeV.

\begin{center}
\begin{table}
\caption{TBEs and excitation energies (MeV) of the lowest $^6$He nucleus states ($T$=1).}
\begin{tabular*}{0.4\textwidth}{ c c c c c c c}
\hline\hline\noalign{\smallskip}
$J^{\pi}$   &  $E_{exp.}$ \cite{exp7he}   & $E^{lim.}_{th.}$ & $E^{extr.}_{th.}$ &  $E^{*}_{exp.}$ & $E^{^*lim.}_{th.}$ & $E^{^*extr.}_{th.}$ \\

\hline\hline\noalign{\smallskip} 
 0$^{+}_1$  & 29.269 &29.239 &29.397&0& 0 &0\\
 
\noalign{\smallskip}

2$^{+}_1$   &27.472& 27.199 & 27.533&1.797& 2.040& 1.864 \\

\noalign{\smallskip}
2$^{+}_2$  &23.7&24.161 & 25.856 & 5.6 &5.087 & 3.541\\

\hline\noalign{\smallskip}
\end{tabular*}

\end{table}
\end{center}

An even more expressive pattern is observed for the lower resonance level 3/2$^{-}_1$ of the $^7$He nucleus and the higher resonance 5/2$^{-}_1$, for which experimental data have been obtained so far. For the former one the calculations on the limited basis underestimate TBEs by 200 keV; for the latter case, this underestimation exceeds 1 MeV. In contrast, the extrapolated results are in good agreement with  the measurement ones. The differences between the results obtained on the limited basis and the extrapolation results for other levels that have not yet been discovered also exceed 1 MeV. 

In Tab. II and its caption, the energies of neutron decay of $^7$He nucleus into the $^6$He$_{gs} $ channels presented in published papers are also given, both calculated and obtained in experiments. The experimental data are better described by the GSM \cite{gamov3}, although the energy higher lying level is reproduced rather well in all approaches. Our data correlate fairly well with the data obtained within the framework of the NCSMC \cite{ncsmc1}. It is interesting to compare the resonance energies which are calculated in the current work with the same values obtained in the framework of approach SS-HORSE \cite{sshorse2}, since this approach, first, is also based on NCSM and, second, this work also uses the NN-potential of Daejeon16. This comparison shows that the use of the extrapolation procedure lowers the values of the resonance energies more than the simultaneous use of a higher value of the cut-off parameter $N^{*max}_{tot}$ and the inclusion of the continuous spectrum in the calculations performed by SS-HORSE method.

\begin{center}
\begin{table}
\caption{TBEs (MeV) and resonance energies (keV) of $^7$He nucleus states ($T$=3/2). a) -- $E^{exp.}_n$=430 keV \cite{exp1}; b) -- $E^{exp.}_n$=3360 keV \cite{exp7he}.}
\begin{tabular*}{0.47\textwidth}{ c c c c c c c c}
\hline\hline\noalign{\smallskip}
$J^{\pi}$   &  $E_{exp.}$ \cite{exp7he}  & $E^{lim.}_{th.}$ & $E^{extr.}_{th.}$&  $E^{extr.}_n$&$E_n$\cite{ncsmc1}  &$E_n$ \cite{gamov3}& $E_n$\cite{sshorse2}\\

\hline\hline\noalign{\smallskip} 
 3/2$^{-}_1$  & 28.83$^a$ & 28.625 & 28.850& 547& 710 & 390& 240\\
 
\noalign{\smallskip}

 1/2$^{+}_1$   & --- & 26.057 & 27.701 & 1696 & --- & --- & ---\\

\noalign{\smallskip}

  1/2$^{-}_1$  & --- & 25.864 & 27.079 & 2318 & 2390 & --- & 2700\\
 
\noalign{\smallskip}
 5/2$^{-}_1$  & 25.91$^b$ & 24.743 & 25.960 & 3437& 3130 & 3470 & 3630\\
 
\noalign{\smallskip}
 3/2$^{+}_1$  & --- & 24.115 & 25.905 & 3492 & --- & --- & 4100\\
 
\noalign{\smallskip}
 5/2$^{+}_1$  &---&23.937 &25.833 &3564 & --- & --- & 4200\\
 
\noalign{\smallskip}
 3/2$^{-}_2$  & --- & 23.966 & 25.455 & 3921 & --- & --- & ---\\

\noalign{\smallskip}

\hline\noalign{\smallskip}
\end{tabular*}

\end{table}
\end{center}

The results of our calculations of the decay widths of  $^7$He nucleus resonances are presented in Tab. III. A wide list of decay channels was analysed, including decays into channels in which fragment $^6$He is in 2$^{+}_1$ or 2$^{+}_2$ excited state. In some cases the partial widths of decay into channels containing the former state are dominant, as can be seen from Tab. III. Because of the gigantic decay width of the latter state, the results of calculations of the characteristics of the channels in which it is contained are physically of little interest. By including the channels containing this state in the analysis, we, first, demonstrate the capabilities of the developed method and, second, we follow the tradition laid down in  Ref. \cite{ncsmc1}.

\begin{center}
\begin{table*}
\caption{Resonance energies, decay widths of open channels (keV), and channel spectroscopic factors of $^7$He nucleus states. Numeric subscript denotes the value $N^{*max}_{tot}$.\\
  a) -- for the experimental value of 3/2$^{-}_1$ state resonance energy 430 keV computed value $\Gamma$ =250 keV; b) -- there are other experimental results: 2.0 MeV \cite{exp2} and 10.0 MeV \cite{exp3}.}
\begin{tabular*}{0.9\textwidth}{ c c c c c c c c c c c c c}
\hline\hline\noalign{\smallskip}
$J^{\pi}$ ($^7$He)   & $J^{\pi}$ ($^6$He) & $E^{lim.}_n$ & $E^{extr.}_n$  & l(S) & SF & $\Gamma^{lim.}$ &$\Gamma^{extr.}_{11}$ & $\Gamma^{extr.}_{13}$ &$\Gamma_{tot}$  \cite{ncsmc1}  &  $\Gamma_{tot}$ \cite{gamov3}& $\Gamma_{tot}$ \cite{sshorse2} & $\Gamma^ {exp.}_{tot}$\\

\hline\hline\noalign{\smallskip}  

3/2$^{-}_1$ & 0$^+_1$ & 614 & 547 & 1(1/2)& 0.730 & 387& 336 & 334$^a$  & 300 & 178 & 110 & 182\\ 

\hline\noalign{\smallskip}

1/2$^{+}_1$ & 0$^+_1$ & 3184 & 1696 & 0(1/2) & 0.844 & 3670 & 2670& 2670 & --- & --- & --- & ---\\ 

\hline\noalign{\smallskip}

  1/2$^{-}_1$ & 0$^+_1$ & 3375 & 2318 & 1(1/2) & 0.814 & 2440 & 1940& 1850 & 2890 & --- & 4300 & 750$^b$ \\
 \cline{2-13} 
& 2$^+_1$  & 1335  & 454 & \begin{tabular}{c} 1(3/2) \\ 3(5/2) \end{tabular} &  \begin{tabular}{c} 0.509 \\ 0.21$\cdot 10^{-3}$  \end{tabular} & \begin{tabular}{c} 812\\-----  \end{tabular} & \begin{tabular}{c}  253\\ 0.47 eV \end{tabular} & \begin{tabular}{c} 221 \\0.46 eV  \end{tabular}  & --- & ---& --- & --- \\

\hline\noalign{\smallskip}

 5/2$^{-}_1$ & 0$^+_1$ & 4496 & 3437 & 3(1/2) & $0.37\cdot 10^{-3}$ & 110 eV& 56 eV & 52 eV & --- & --- & --- & ---\\ 
 \cline{2-13}  
& 2$^+_1$  & 2606 & 1573 & \begin{tabular}{c} 1(3/2)\\ 1(5/2) \end{tabular} &  \begin{tabular}{c} 0.420  \\ 0.758  \end{tabular} & \begin{tabular}{c} 1366 \\ 1783 \end{tabular} & \begin{tabular}{c} 980\\1200  \end{tabular} & \begin{tabular}{c} 881 \\  1060   \end{tabular}  & 1070 & 2300 & 1360 & 1990 \\

\hline\noalign{\smallskip}

 3/2$^{+}_1$ & 0$^+_1$ & 5124 & 3492 & 2(1/2) & 0.041 & 125 & 84.4 & 83.4 & --- & --- & --- & --- \\ 
 \cline{2-13}  
&2$^+_1$  & 3084 & 1628 & \begin{tabular}{c} 0(3/2) \\2(3/2) \end{tabular} &  \begin{tabular}{c} 0.752 \\ 0.051  \end{tabular} & \begin{tabular}{c} 3430 \\ 58  \end{tabular} & \begin{tabular}{c} 2590 \\22.3  \end{tabular} & \begin{tabular}{c}  2490 \\ 17.9 \end{tabular}  & --- & --- & 4400& ---\\
 
\hline\noalign{\smallskip}

 5/2$^{+}_1$ & 0$^+_1$ & 5302 & 3564 & 2(1/2) & 0.126 & 382 & 285 & 258 & --- & --- & --- & --- \\ 
 \cline{2-13}  
& 2$^+_1$  & 3262 & 1700 & \begin{tabular}{c} 0(5/2) \\ 2(5/2)\end{tabular} &  \begin{tabular}{c} 0.704  \\ 0.019  \end{tabular} & \begin{tabular}{c} 3100 \\ 29  \end{tabular} & \begin{tabular}{c} 2210 \\13.7 \end{tabular} & \begin{tabular}{c} 2240 \\ 11.1  \end{tabular}  & --- & --- &  5000 & --- \\
 \cline{2-13} 
& 2$^+_2$  & 224  & 23 & \begin{tabular}{c} 0(5/2) \\ 2(3/2) \end{tabular} &  \begin{tabular}{c} 0.001 \\ 0.012  \end{tabular} & \begin{tabular}{c} 9.3 \\ 38.6 eV \end{tabular} & \begin{tabular}{c} 3.33 \\0.15 eV  \end{tabular} & \begin{tabular}{c}  2.98 \\  0.14 eV \end{tabular}  & --- & --- & --- & --- \\

\hline\noalign{\smallskip}

 3/2$^{-}_2$ & 0$^+_1$ & 5273 & 3921 & 1(1/2) & 0.069 & 276 & 249 & 229 & --- & --- & --- & --- \\ 
  \cline{2-13} 
& 2$^+_1$  & 3233 & 2057 & \begin{tabular}{c} 1(3/2) \\ 1(5/2) \end{tabular} &  \begin{tabular}{c} 0.190  \\0.470 \end{tabular} & \begin{tabular}{c} 556 \\ 1486 \end{tabular} & \begin{tabular}{c}  408 \\ 1150  \end{tabular} & \begin{tabular}{c}  399 \\ 1060 \end{tabular}  & --- & --- & --- & --- \\
 \cline{2-13} 
& 2$^+_2$  & 196 & 380 & 1(3/2) & 0.268 & 44& 126 & 108 & --- & --- & --- & --- \\

\hline\noalign{\smallskip}
\end{tabular*}

\end{table*}
\end{center}

It is natural to start the discussion by assessing the accuracy of the method used. Comparison of the results contained in the 3rd and 4th and, respectively, in the 7th and 9th columns of Tab. III, demonstrates a strong dependence of the decay width of a certain state on its energy. Roughly assessing the tendency, one may say that the relative change in the decay energy leads to a proportional change in the width. From this point of view, the relative error in computing the decay energy value of a particular state is reproduced when calculating its width. The relative decay energy errors of ab initio computations on a limited basis for highly excited levels, are quite large because decay energy values themselves are the difference between TBEs. Therefore, the introduction of approaches that refine the energy values, in particular the extrapolation method, is extremely important.

The size of the used basis affects the behavior of the CFF curves, which also exerts some influence on the magnitude of the decay width, however this influence is somewhat weaker in comparison with the effect produced by the change in the decay energy. The scale of the change in the decay widths with a change in the size of the basis from $N^{*max}_{tot}$ = 11 to $N^{*max}_{tot}$= 13 is illustrated by columns 8 and 9 of Tab. III. The computation results show that convergence has actually been achieved for a significant part of the examples. Changes in decay widths for other examples are in the region of 10\%. Only in two cases does this change exceed 20\%. There is no doubt that this accuracy is sufficient for planning experiments aimed at studying the spectra of exotic nuclei.

Comparison of the obtained results with experimental data leads to the following conclusions. TBEs of the levels of $^7$He nucleus, values of which were measured, are very well reproduced in the framework of the calculations. Therefore the first source of the discrepancy between the calculated and measured values of the decay width  of the ground state of the $^7$He nucleus is the overestimation of the binding energy of the $^6$He nucleus. This overestimation, evidently, has nothing to do with the choice of the parameters of the NCSM basis and, very likely, with the peculiarities of the extrapolation procedure. Its causes are, most likely, the properties of the Daejeon16 potential.  The second source of the 30\% overestimation of the decay width, revealed by its calculation using the experimental decay energy, is not clear, since all the conditions listed above for correct matching in this example are satisfied, and the convergence of the result with respect to $N^{*max}_{tot}$ is achieved. In principle, it can be assumed that this discrepancy is generated by the procedure of extracting the width from the experiment. Indeed, there are a variety of procedures used by evaluators to determine the decay energy and width, and these procedures contain various variation parameters. Thirty percent differences in the results of processing the same experiments can be found in the same databases (see, for example, \cite{exp7he}). On the other hand the GSM calculations \cite{gamov3} are in a good agreement with the experimental data, while other theoretical methods show overestimation or underestimation of 3/2$^-_1$ resonance width. So, in our opinion, the issue remains open.

The calculated values of the decay energy of state 5/2$^{-}_1$ which is equal to 3437 keV and its total width 1941 keV are in a very good agreement with the experimental data. A good agreement was also achieved in the calculations based on the GSM \cite{gamov3}. 

The differences between the predicted values of the energies and decay widths of the $^7$He nucleus states obtained by us and the results of other authors are large enough. That makes future experiments exciting. 

\begin{center}
\begin{table}
\caption{Channel spectroscopic factors of closed and strongly suppressed decay channels of $^7$He nucleus states. Symbol * indicates closed channels}
\begin{tabular*}{0.29\textwidth}{ c c c c}
\hline\hline\noalign{\smallskip}
$J^{\pi}$ ($^7$He)   & $J^{\pi}$ ($^6$He)   & l(S) & SF\\

\hline\hline\noalign{\smallskip}  

3/2$^{-}_1$ 
&2$^{+*}_1$   &  \begin{tabular}{c} 1(3/2) \\ 1(5/2) \\3(3/2) \\ 3(5/2)\end{tabular} &  \begin{tabular}{c} 0.691 \\0.254 \\ 0.0006\\ 0.0046  \end{tabular} \\
 \cline{2-4} \noalign{\smallskip}
&2$^{+*}_2$   &  \begin{tabular}{c}1(3/2) \\ 1(5/2) \\3(3/2) \\ 3(5/2) \end{tabular} &  \begin{tabular}{c} 0.338 \\0.0015 \\ 0.0003\\ 0.00243 \end{tabular} \\

\hline\noalign{\smallskip}  

1/2$^{+}_1$ 
&2$^{+*}_1$   &  \begin{tabular}{c} 2(3/2) \\ 2(5/2) \end{tabular} &  \begin{tabular}{c} 0.0258\\ 0.0198 \end{tabular} \\
 \cline{2-4} 
&2$^{+*}_2$   &  \begin{tabular}{c} 2(3/2) \\ 2(5/2) \end{tabular} &  \begin{tabular}{c}  0.0117\\0.0017  \end{tabular} \\

\hline\noalign{\smallskip}  

1/2$^{-}_1$  
&2$^{+*}_2$   &  \begin{tabular}{c}1(3/2) \\ 3(5/2)  \end{tabular} &  \begin{tabular}{c} 0.562 \\ 0.0004 \end{tabular} \\

\hline\noalign{\smallskip}  

5/2$^{-}_1$ 
&2$^+_1$   &  \begin{tabular}{c} 3(3/2) \\ 3(5/2)\end{tabular} &  \begin{tabular}{c}0.0098 \\0.0036 \end{tabular} \\
 \cline{2-4} 
&2$^{+*}_2$   &  \begin{tabular}{c}1(3/2) \\ 1(5/2) \\3(3/2) \\ 3(5/2) \end{tabular} &  \begin{tabular}{c} 0.641 \\ 0.604 \\0.0008 \\ 0.0001 \end{tabular} \\

\hline\noalign{\smallskip}  

3/2$^{+}_1$  
&2$^+_1$   &  2(5/2)  &  0.0052  \\
 \cline{2-4} 
&2$^{+*}_2$   &  \begin{tabular}{c}  0(3/2) \\ 2(3/2) \\2(5/2)\end{tabular} &  \begin{tabular}{c} 0.124 \\ 0.014 \\0.0005  \end{tabular} \\

\hline\noalign{\smallskip}  

5/2$^{+}_1$ 
&2$^+_1$   &  \begin{tabular}{c} 2(3/2) \\ 4(3/2) \\4(5/2) \end{tabular} &  \begin{tabular}{c} 0.0052 \\0.0006 \\0.0000 \end{tabular} \\
 \cline{2-4} 
&2$^+_2$   &  \begin{tabular}{c} 2(5/2) \\ 4(3/2) \\4(5/2) \end{tabular} &  \begin{tabular}{c}  0.0079\\  0.0001\\ 0.0002 \end{tabular} \\

\hline\noalign{\smallskip}  

3/2$^{-}_2$  
&2$^+_1$  &  \begin{tabular}{c} 1(5/2) \\ 3(3/2) \\ 3(5/2) \end{tabular} &  \begin{tabular}{c} 0.0025 \\0.0064  \\0.0002\end{tabular} \\
 \cline{2-4} 
&2$^+_2$  &  \begin{tabular}{c} 1(5/2) \\3(3/2) \\ 3(5/2)\end{tabular} &  \begin{tabular}{c} 0.0025  \\0.0064 \\0.0002  \end{tabular} \\

\hline\noalign{\smallskip}
\end{tabular*}

\end{table}
\end{center}
 
Approaching the conclusion, it should be emphasized that one of the main goals of this work is to support experiments aimed at studying spectrum of $^7$He that are carried out or planned. We believe that the data presented in Tab. III can be used as preliminary for the analysis of the decay properties of these states. For the same purposes, we consider it useful to supplement the tabular data presented above with one more table. Tab. IV contains SFs of many channels not discussed above. The data presented in the table indicate that in order to analyse the spectrum of the nucleus, one can restrict oneself to the channels, the parameters of which are presented in tab. III, since the other channels are either closed or their SFs are small.

\section{Conclusions}

The main results of this work are the following.

\noindent
I. Based on the successful application of the Cluster Channel Orthogonalized Functions Method for describing  $^7$Li and $^8$Be nuclei, total binding energies (TBEs), excitation energies, as well as decay energies and widths of the levels of $^7$He nucleus are calculated in this NCSM-based ab initio scheme. Extrapolation A5 method is involved for obtaining TBEs with better precision.

\noindent
II. The approach used allows one to solve multichannel problems, which made it possible for the first time to calculate in an ab initio scheme not only the total but the partial widths of decay of the $^7$He nucleus into many exit channels.

\noindent
III. The studies carried out demonstrate good prospects of the method used for the theoretical study of neutron-rich nuclei, in particular, for predicting the results of planned experiments.

\noindent
IV. The prospects for using this approach in the field of studies of the interaction of neutrons with light nuclei seem to be even wider.

\section{Acknowledgments}

 We are grateful to A. M. Shirokov, A. I. Mazur and I. A. Mazur for fruitful discussions and providing us the Daejeon16 NN-potential matrixes, as well as   to C. W. Johnson for supporting our efforts to introduce high-performance code Bigstick for our NCSM calculations.

\end{document}